\documentclass[a4paper,11pt]{ijamas}

\usepackage{graphicx}

\begin{document}


\title{Complexity of multidimensional hydrogenic systems}

\author{\textbf{Jes\'us S. Dehesa$^{1,2}$, Sheila L\'opez-Rosa$^{1,2}$, Pablo S\'anchez-Moreno$^{2,3}$ and Rafael J. Y\'a\~nez$^{2,3}$}}

\date{$^1$Departamento de F\'isica At\'omica, Molecular y Nuclear\\
  Universidad de Granada\\
 $^2$Instituto ''Carlos I`` de F\'isica Te\'orica y Computacional\\
 Universidad de Granada\\
$^3$Departamento de Matem\'atica Aplicada\\
  Universidad de Granada\\
}

\maketitle


\begin{abstract}
\noindent \emph{The Cram\'er-Rao, Fisher-Shannon and LMC shape complexity measures have
been recently shown to play a relevant role to study the internal disorder
of finite many-body systems (e.g., atoms, molecules, nuclei). They
highlight amongst the bunch of complexities appeared in the literature
by their mathematical simplicity and transparent physical interpretation.
They are composed by two spreading measures (variance,
Fisher information and Shannon entropies) of the single-particle
probability density of the systems. Here we discuss the
physico-mathematical knowledge of these two-component complexities for
hydrogenic systems with standard ($D = 3$) and non-standard
dimensionalities. These systems include not only the hydrogen atom and its
isoelectronic series but also a large diversity of quantum systems such
as, e.g. exotic atoms, antimatter atoms, some qubits systems and Rydberg
atoms, ions and molecules.
}

\medskip


\medskip


\end{abstract}


\section{Introduction}
\label{sec:section1}

In this work we describe the physico-mathematical bases of the internal disorder
of the hydrogenic systems, which is manifest in the non-uniformity of their quantum-mechanical single-particle probability density $\displaystyle{\rho(\vec{r})}$ and in the so distinctive hierarchy of stationary physical states, by means of the following two-component information-theoretic composite measures: the Cram\'er-Rao, Fisher-Shannon and LMC shape complexities. These quantities, which measure how easily the systems may be modelled, are much closer to the everyday notion of complexity than many other abstract notions of structure and complexity \cite{rescher_98,goldreich_08,bonchev_03}.\\

THe $D$-dimensional hydrogenic system (i.e., a negatively-charged particle moving around a positively-charged core which electromagnetically binds it in its orbit) plays a central role in $D$-dimensional quantum physics and chemistry \cite{herschbach_93}. It encompasses not only the hydrogen atom and its isoelectronic series but also the Rydberg atoms and molecules, some exotic and antimatter atoms, and numerous low-dimensional nanostructured entities (quantum wells, wires and dots) and qubits systems. The existence for such a system has been shown for $D\leq 3$ \cite{li:pla07} and suggested for $D>3$ \cite{burgbacher:jmp99}. Moreover, the position and momentum multidimensional hydrogenic wavefunctions are often used as complete orthonormal sets for many-body problems \cite{aquilanti:aqc01,aquilanti:irpc01}.\\

The information-theoretic analysis of the multidimensional hydrogenic system was initiated in 1994 \cite{yanez:pra94} by extending to arbitrary dimensions and to momentum space some previous works on the Shannon entropy \cite{shannon:bst48} of the ground state of the three-dimensional hydrogen atom done by Gadre et al during the nineteen eighties (see e.g. \cite{gadre:pra87}; see also the survey \cite{gadre:aqc91}); see also Chen et al \cite{chen:jmc94}. Therein, we \cite{yanez:pra94} studied the dimensionality dependence of the ground-state hydrogenic sytem and the values of the Shannon entropy of a large class of excited states for the $D$-dimensional hydrogen atom in a numerical way for $D=1,2$ and $3$, and we started to realize the crucial role which the emerging information theory of the classical orthogonal polynomials was going to play for the theoretical determination of the hydrogenic information-theoretic quantities. The state-of-the-art of this theory and its application to the analytical determination of the position and momentum Shannon entropy $\displaystyle{S \left[ \rho \right]}$ of $D$-dimensional single-particle systems subject to central potentials in general and hydrogenic system in particular, was done up until 2001 in \cite{dehesa:jcam01}. Five years later, the Fisher information \cite{fisher:pcps25,frieden_04} of the ground and excited states of the multidimensional hydrogenic system was analytically found \cite{dehesa:jmp06,romera:jmp06} in both position and momentum spaces by two different approaches: one based purely in the algebraic properties of the classical orthogonal polynomials \cite{dehesa:ijqc10} and another one using its connection with the second-order moments $\displaystyle{\left\langle r^\alpha \right\rangle}$ and $\displaystyle{\left\langle p^\alpha \right\rangle}$ \cite{romera:jmp06}. The present knowledge about the single-componet information-theoretic quantities (moments around the origin, logarithmic moments, variance $\displaystyle{V \left[ \rho \right]}$, Shannon entropy  $\displaystyle{S \left[ \rho \right]}$ and Fisher information  $\displaystyle{F \left[ \rho \right]}$) and their associated uncertainty relations has been throughly considered this year in Ref. \cite{dehesa:ijqc10}.\\

Nowadays, the internal order of the multidimensional hydrogenic systems can be analyzed in a deeper and more complete way with the advent of the following two-component information-theoretic measures: the LMC shape complexity \cite{lopezruiz:pla95,catalan:pre02} defined by
\begin{equation}\label{eq:definicion_LMC}
C_{\text{LMC}} \left[ \rho \right] := D \left[ \rho \right] \times \exp \left( S \left[ \rho \right]\right),
\end{equation}
the Fisher-Shannon complexity \cite{romera:jcp04,dembo:ieeetit91} given by
\begin{equation}\label{eq:definicion_FS}
C_{\text{FS}} \left[ \rho \right] := F \left[ \rho \right] \times \frac{1}{2 \pi e} \exp \left(\frac{2}{3} S \left[ \rho \right]\right),
\end{equation}
and the Cram\'er-Rao complexity \cite{angulo:jcp08,cover_91} defined by
\begin{equation}\label{eq:definicion_CR}
C_{\text{CR}} \left[ \rho \right] := F \left[ \rho \right] \times V \left[ \rho \right],
\end{equation}
where the variance $\displaystyle{V \left[ \rho \right] \equiv \left\langle r^2 \right\rangle - \left\langle r \right\rangle^2}$ and the disequilibrium $\displaystyle{D\left[ \rho \right] \equiv \left\langle \rho \right\rangle }$. The symbols $\displaystyle{\left\langle r^\alpha \right\rangle}$ and $\displaystyle{\left\langle \rho^\alpha \right\rangle}$ for the moments-around-the-origin and the frequency or entropic moments, respectively, are defined as
\begin{equation}\label{eq:definicion_ralpha_rho_alpha}
\left\langle r^\alpha \right\rangle := \int r^\alpha \rho(\vec{r}) d\vec{r}, \quad \text{and} \quad \left\langle \rho^\alpha \right\rangle := \int \left[ \rho(\vec{r}) \right]^{\alpha+1} d\vec{r},
\end{equation}
respectively. These three measures of complexity are complementary in the sense that they quantify, according to their information-theoretic composition, different facets of the internal disorder of the system which give rise to the great diversity and complexity of the $D$-dimensional geometries of the configuration shapes of the probability density $\displaystyle{\rho(\vec{r})}$ corresponding to its orbitals characterized by $D$ integer hyperquantum numbers $\displaystyle{\left\lbrace \mu_0, \dots, \mu_{D-1} \right\rbrace }$. The LMC shape complexity has two global ingredients, the disequilibrium and the Shannon entropy power, so that it quantifies the combined balance of the average height and the total extent of the electron distribution. A contrario, the Fisher-Shannon and Cram\'er-Rao quantities have an ingredient with a local property, the Fisher information, which is very sensitive to the electronic oscillations. The Fisher-Shannon complexity quantifies the gradient content jointly with the total extent of $\displaystyle{\rho(\vec{r})}$, while the Cram\'er-Rao complexity measures the gradient content of $\displaystyle{\rho(\vec{r})}$ jointly with its spreading around the centroid. Let us also remark that the three complexity measures are minimum for the two extreme (or least complex) probability distributions, which correspond to perfect order (mathematical represented by a extremely localized density) and maximum disorder (associated to a highly flat distribution). Moreover, they fulfill invariance properties under replication, translation and scaling transformation \cite{catalan:pre02,lopezrosa:pa09}. Finally, let us collect here that these quantities satisfy the following lower bounds
\begin{equation}\label{eq:cotas_complejidad}
C_{\text{LMC}} \left[ \rho \right] \geq 1, \qquad C_{\text{FS}} \left[ \rho \right] \geq D \quad \text{and} \quad C_{\text{CR}} \left[ \rho \right] \geq D^2
\end{equation}
for general $D$-dimensional probability densities \cite{lopezrosa:pa09}.\\

The structure of this paper is the following. First, in Section \ref{sec:section2} we give the physico-mathematical $D$-dimensional methodology to determine the three hydrogenic complexity measures mentioned above. Then, we monographically discuss the LMC shape complexity of the ground and excited circular $D$-dimensional hydrogenic orbitals. Finally, some conclusions and open problems are given.



\section{Hydrogenic complexity methodology}
\label{sec:section2}

In this Section we first give the quantum-mechanical probability density of the $D$-dimensional ($D >1$) hydrogenic system in position and momentum spaces. Then, we describe the theoretical methodology to compute the two-component complexity measures which quantify the spatial spreading of the associated position and momentum electron distributions.\\

The electronic orbitals of the $D$-dimensional quantum-mechanical problem for the Coulumb potential $\displaystyle{V_D(r)=-\frac{Z}{r}}$, $\displaystyle{r=\left| \vec{r} \right|}$, have been shown (see e.g., \cite{nieto:ajp79,yanez:pra94,dehesa:ijqc10}) to be described by the probability density
\begin{equation}\label{eq:rho_Dens}
\rho_{n,l,\left\lbrace \mu \right\rbrace} (\vec{r}) = \frac{\lambda^{-D}}{2 \eta} \frac{\omega_{2L+1}(\hat{r})}{{\hat{r}}^{D-2}} \left[ {\tilde{\cal{L}}}_{\eta-L-1}^{2L+1}(\hat{r}) \right]^2  \left|{\cal{Y}}_{l,\left\lbrace \mu \right\rbrace}
 \left( \Omega_{D}\right)  \right|^2 ,
\end{equation}
in configuration space, where the postion vector $\displaystyle{\vec{r}=(r, \theta_1,\cdots,\theta_{D-1})}$ in polar hyperspherical coordinates, $\displaystyle{(n,l,\left\lbrace \mu \right\rbrace)=\left(n,l \equiv \mu_1,\cdots, \mu_{D-1} \right)}$ are the associated hyperquantum numbers which may have the values
\begin{equation}\label{eq:quantum_numbers}
n=1,2,3,\dots ; \qquad l=0,1,2,\dots,n-1; \qquad l \geq \mu_2 \geq \dots \geq \mu_{D-1}=\left| m \right| \geq 0
\end{equation}
and
\begin{equation}\label{eq:eta_L_r}
L \equiv l + \frac{D-3}{2}, \qquad \hat{r}=\frac{r}{\lambda}, \qquad \lambda=\frac{\eta}{2Z} \quad \text{with} \quad \eta=n+\frac{D-3}{2}
\end{equation}

The symbol $\displaystyle{{\tilde{\cal{L}}}_{k}^{\alpha}(x)}$ denotes the Laguerre polynomials of degree $k$ and parameter $\alpha$, orthogonal with respect to the weight function $\displaystyle{\omega_{\alpha}(x)=x^{\alpha}e^{-x}}$. Moreover, the angular part $\displaystyle{{\cal{Y}}_{l,\left\lbrace \mu \right\rbrace}
 \left( \Omega_{D}\right) }$ denotes the hyperspherical harmonics \cite{avery_00,yanez:pra94,dehesa:ijqc10} so that
\begin{equation}\label{eq:hiperesfericos}
{\cal{Y}}_{l,\{\mu\}}(\Omega_{D-1})=\frac{1}{\sqrt{2 \pi}} 
e^{im\varphi} \prod^{D-2}_{j=1} {\tilde{C}}_{\mu_{j}-\mu_{j+1}}^{\alpha_j+\mu_{j+1}}
(\cos \theta_j) \left( \sin \theta_j\right)^{\mu_{j+1}},
\end{equation}
with $\alpha_j= \frac{1}{2} (D-j-1)$ and ${\tilde{C}}^{\lambda}_{k}(x)$ denotes the 
orthonormal Gegenbauer polynomials of degree $k$ and parameter $\lambda$.\\

Similarly, the momentum electronic orbitals of the system are given (see e.g. \cite{avery_00,avery:jpc93,dehesa:ijqc10}) as
\begin{equation}\label{eq:gamma_Dens}
\gamma(\vec{p})= \left(  \frac{\eta} {Z}\right) ^{D} (1+y)^{3} \left(\frac{1+y}{1-y} \right)^{\frac{D-2}{2}}
 {\omega^{*}}_{L+1} (y) \left[ {\tilde{C}}^{L+1}_{\eta-L-1}(y) \right]^2 \left[ {\cal{Y}}_{l\{\mu\}}(\Omega^{'}_{D-1})\right]^2,
\end{equation}
with $\displaystyle{\vec{p}=(p, \theta^{'}_{1},\cdots,\theta^{'}_{D-1}) \equiv \left(p,\Omega^{'}_{D-1} \right)}$ and the notation
\begin{equation}\label{eq:y_p_omegaC}
y=\frac{1-\eta^2 \tilde{p}^2 }{1+\eta^2 \tilde{p}^2}, \quad \tilde{p}=\frac{p}{Z} \quad \text{and} \quad \omega^{*}_{\alpha}(x)=(1-x^2)^{\alpha-\frac{1}{2}}.
\end{equation}

The corresponding stationary states of the system are known to have the energies $\displaystyle{E_\eta=- \frac{Z^2}{\eta^2}}$, where $\eta$ denotes the grand quantum number defined in Eq. (\ref{eq:eta_L_r}).\\

\textbf{LMC shape complexity}\\

Let us now show the theoretical methodology to determine the LMC shape complexity \cite{lopezruiz:pla95,catalan:pre02} defined by Eq. (\ref{eq:definicion_LMC}) which has been recently and successfully used for different purposes \cite{sanudo:pla08,romera:irp09,lopezrosa:pa09b}. This quantity has two ingredients: the disequilibrium and the Shannon entropy. The disequilibrium turns out \cite{lopezrosa:pa09b} to be
\begin{equation}\label{eq:Drho1}
D \left[ \rho \right] = \int \rho(\vec{r})^2 d\vec{r} = \frac{2^{D-2}}{\eta^{D+2}} Z^D K_1
 \left( D,\eta,L \right) K_2 \left( l, \left\lbrace \mu \right\rbrace \right),
\end{equation}
and
\begin{equation}\label{eq:Dgamma1}
D\left[ \gamma \right] = \int \gamma^2 \left(\vec{p} \right) d \vec{p}=
 \frac{2^{4L+8} \eta^{D} }{Z^D} K_3 \left( D,\eta,L \right) K_2 
\left( l, \left\lbrace \mu \right\rbrace \right),
\end{equation}
for the position and momentum electron densities given by Eqs. (\ref{eq:rho_Dens}) and (\ref{eq:gamma_Dens}) respectively. The symbols $\displaystyle{K_i}$ ($i=1,2$ and $3$) denote the following functionals of Laguerre polynomials
\begin{equation}\label{eq:int_K1}
K_1 \left( D,\eta,L \right) = \int_{0}^{\infty} x^{-D-5} \left\lbrace \omega_{2L+1} (x) 
\left[ {\tilde{{\cal{L}}}}^{2L+1}_{\eta-L-1} (x) \right]^2  \right\rbrace^2 dx,
\end{equation}
of hyperspherical harmonics
\begin{equation}\label{eq:int_K2}
K_2 \left( l, \left\lbrace \mu \right\rbrace \right) = \int_{\Omega} \left| {\cal{Y}}_{l\{\mu\}}
\left(\Omega_{D-1}\right) \right|^4 d\Omega_{D-1},
\end{equation}
and of Gegenbauer polynomials
\begin{equation}\label{eq:int_K3}
K_3 \left( D,\eta,L \right) = \int_{0}^{\infty} \frac{y^{4l+D-1}}{(1+y^2)^{4L+8}} 
\left[ {\tilde{C}}^{L+1}_{\eta-L-1} \left(\frac{1-y^2}{1+y^2}\right) \right]^4  dy.
\end{equation}

On the other hand, the total spread of $\displaystyle{\rho(\vec{r})}$ given by Eq. (\ref{eq:rho_Dens}) is best quantified by the position Shannon entropy
\begin{align}\label{eq:shannon_rho}
S \left[ \rho \right]:&= - \int \rho(\vec{r}) \ln \rho(\vec{r}) d\vec{r} \nonumber\\
&= A(n,l,D) + \frac{1}{2\eta}  E_1\left[ \tilde{{\cal{L}}}^{2L+1}_{\eta-L-1} \right]-D \ln Z+ B (l,\left\lbrace \mu \right\rbrace,d)+\sum^{d-2}_{j=1} E_0 
\left[ \tilde{C}^{\alpha_j+\mu_{j+1}}_{\mu_j-\mu_{j+1}} \right],
\end{align}
where
\begin{equation}\label{eq:coefA}
A(n,l,D)= -2l \left[\frac{2\eta-2L-1}{2 \eta} +\psi (\eta+L+1) \right]+\frac{ 3 \eta^2-L(L+1)}{\eta}
+ -\ln \left[ \frac{2^{D-1}}{\eta^{D+1}} \right],
\end{equation}
and
\begin{align}\nonumber
B (l,\left\lbrace \mu \right\rbrace,D)&= \ln 2\pi -2 \sum^{D-2}_{j=1} \mu_{j+1} \\ \label{eq:coefB}
&\quad \times \left[\psi(2\alpha_j+\mu_j+\mu_{j+1})-\psi(\alpha_j+\mu_j)-\ln 2- 
\frac{1}{2 (\alpha_j+\mu_j)}\right],
\end{align}
and
\begin{equation}\label{eq:int_Ei}
E_i \left[p_n \right]=-\int_{0}^{\infty} x^{i} \omega(x) p^{2}_{n} (x) \ln p^{2}_{n}(x) dx, \quad i=0,1
\end{equation}
is the Shannon entropic integral of the polynomials $p_n(x)$ orthogonal with respect to the weight function $\omega(x)$.\\

Similarly, the total spread of $\displaystyle{\gamma(\vec{p}}$ given by Eq. (\ref{eq:gamma_Dens}) can be calculated by means of the momentum Shannon entropy
\begin{align}\label{eq:shannon_gamma}
S \left[ \gamma \right]:&= - \int \gamma(\vec{p}) \ln \gamma(\vec{p}) d\vec{p} \nonumber\\
&= F(n,l,D) +  E_0\left[ \tilde{{\cal{C}}}^{L+1}_{\eta-L-1} \right]+D \ln Z+ B (l,\left\lbrace \mu \right\rbrace,D)+\sum^{D-2}_{j=1} E_0 
\left[ \tilde{C}^{\alpha_j+\mu_{j+1}}_{\mu_j-\mu_{j+1}} \right],
\end{align}
where
\begin{align}\label{eq:coef_F}
F\left(n,l,D \right)&=-\ln \frac{\eta^D}{2^{2L+4}}-(2L+4) \left[ \psi(\eta+L+1)-\psi (\eta) \right]
\nonumber \\
& \quad +\frac{L+2}{\eta}-(D+1) \left[1-\frac{2 \eta (2L+1)}{4\eta^2-1} \right].
\end{align}

Then, the combined use of Eqs. (\ref{eq:definicion_LMC}), (\ref{eq:rho_Dens}), (\ref{eq:Drho1}) and (\ref{eq:shannon_rho}) yields the following value
\begin{align}\label{eq:Clmc_posiciones}
C_{\text{LMC}} \left[ \rho \right]:& = D\left[ \rho \right] \times \exp \left( S \left[ \rho \right]\right) \nonumber \\
& = \frac{2^{D-2}}{\eta^{D+2}} K_1 \left(D, \eta, L \right)
 K_2 \left( L, \left\lbrace \mu \right\rbrace \right) \nonumber \\
& \quad \times \exp \left[ A(n,l,D)+  B (l,\left\lbrace \mu \right\rbrace,D) + \frac{1}{2 \eta} E_1 \left[{\tilde{L}}_{\eta-L-1}^{2 L+1} \right] + \sum^{D-2}_{j=1} E_0 
\left[ \tilde{C}^{\alpha_j+\mu_{j+1}}_{\mu_j-\mu_{j+1}} \right]
\right],
\end{align}
for the LMC shape complexity in position space. Similarly, the combined use of Eqs. (\ref{eq:definicion_LMC}), (\ref{eq:gamma_Dens}), (\ref{eq:Dgamma1}) and (\ref{eq:shannon_gamma}) yields the value
\begin{align}\label{eq:Clmc_momentos}
C_{\text{LMC}} \left[ \gamma \right]:& = D\left[ \gamma \right] \times \exp \left( S \left[ \gamma \right]\right) \nonumber \\
& = 2^{4L+8} \eta^{D} K_3 \left(D, \eta, L \right)
 K_2 \left( L, \left\lbrace \mu \right\rbrace \right) \nonumber \\
& \quad \times \exp \left[ F(n,l,D)+  B (l,\left\lbrace \mu \right\rbrace,D) + E_0 \left[{\tilde{C}}_{\eta-L-1}^{L+1} \right] + \sum^{D-2}_{j=1} E_0 
\left[ \tilde{C}^{\alpha_j+\mu_{j+1}}_{\mu_j-\mu_{j+1}} \right]
\right],
\end{align}
for the LMC shape complexity in momentum space. It is worth noting that LMC complexities in the two reciprocal spaces do not depend on the nuclear charge $Z$ of the system. In fact, this is true not only for the Coulomb potential but also for any homogeneous potential \cite{patil:jpb07}. Moreover, the position LMC complexity depend on some functionals of Laguerre and Gegenbauer polynomials, while for the computation of the momentum LMC complexity only functionals of Gegenbauer polynomials are involved. To understand this, we should keep in mind (i) that the hyperspherical harmonics are closely connected with the Gegenbauer polynomials as given by Eq. (\ref{eq:hiperesfericos}), and (ii) the momentum electron density (\ref{eq:gamma_Dens}) of the system is fully controlled by Gegenbauer polynomials.\\

\textbf{Fisher-Shannon complexity}\\

According to Eq. (\ref{eq:definicion_FS}), the ingredients of the Fisher-Shannon complexity are the Fisher information and the Shannon entropy power duly modified. The Fisher information of the $D$-dimensional hydrogenic system has been recently shown to have the value 
\begin{equation}\label{eq:fisher_rho}
F \left[\rho \right]: = \int \frac{ \left| \vec{\nabla}_D \rho (\vec{r}) \right|^2}{\rho(\vec{r})} d\vec{r} = \frac{4Z^2}{\eta^3} \left[\eta- \left| m \right| \right]  
\end{equation}
in position space \cite{dehesa:jmp06}, and
\begin{equation}\label{eq:fisher_gamma}
F \left[\gamma \right]: = \int \frac{ \left| \vec{\nabla}_D \gamma (\vec{p}) \right|^2}{\gamma(\vec{p})} d\vec{p} = \frac{2\eta^2}{Z^2} \left[ 5 \eta^2-3L(L+1) -\left| m \right|(8\eta-6L-3)+1\right]
\end{equation}
in momentum space \cite{dehesa:jmp06}. Then, from Eqs. (\ref{eq:definicion_FS}), (\ref{eq:shannon_rho}) and (\ref{eq:fisher_rho}) we can write down the expression
\begin{equation}\label{eq:Cfs_posiciones}
C_{\text{FS}} \left[ \rho \right]= \frac{2}{\pi e \eta^3} \left[\eta- \left| m \right| \right] \times \exp \left( \frac{2}{D} T\left[ \rho \right]  \right)
\end{equation}
for the position Fisher-Shannon complexity of the system, where
\begin{equation}\label{eq:coefTrho}
T\left[ \rho \right]= A(n,l,D) + B (l,\left\lbrace \mu \right\rbrace,D) + \frac{1}{2\eta}  E_1\left[ \tilde{{\cal{L}}}^{2L+1}_{\eta-L-1} \right]+\sum^{D-2}_{j=1} E_0 
\left[ \tilde{C}^{\alpha_j+\mu_{j+1}}_{\mu_j-\mu_{j+1}} \right]
\end{equation}

Similarly, from Eqs.(\ref{eq:definicion_FS}), (\ref{eq:shannon_gamma}) and (\ref{eq:fisher_gamma}) one finds the expression
\begin{equation}\label{eq:Cfs_momentos}
C_{\text{FS}} \left[ \gamma \right]= \frac{\eta^2}{\pi e}  \left[ 5 \eta^2-3L(L+1) -\left| m \right|(8\eta-6L-3)+1\right] \times \exp \left( \frac{2}{D} T\left[ \gamma \right]  \right)
\end{equation}
for the momentum Fisher-Shannon complexity of the system, where
\begin{equation}\label{eq:coefTgamma}
T\left[ \gamma \right]= F(n,l,D) + B (l,\left\lbrace \mu \right\rbrace,D) +  E_0\left[ \tilde{{\cal{C}}}^{L+1}_{\eta-L-1} \right]+\sum^{D-2}_{j=1} E_0 
\left[ \tilde{C}^{\alpha_j+\mu_{j+1}}_{\mu_j-\mu_{j+1}} \right]
\end{equation}

In Eqs. (\ref{eq:Cfs_posiciones}) and (\ref{eq:Cfs_momentos}) we note that the Fisher-Shannon complexity of the $D$-dimensional hydrogenic system do not depend either on the Coulomb strength, i.e., on the nuclear charge $Z$.\\

\textbf{Cram\'er-Rao complexity}\\

Let us now compute the Cram\'er-Rao complexity of the system which, according to its definition (\ref{eq:definicion_CR}), has two ingredients: the Fisher information and the variance of the density. The former one is given by Eqs. (\ref{eq:fisher_rho}) and (\ref{eq:fisher_gamma}) in position and momentum spaces, respectively. The variance has been already shown to have the value
\begin{equation}\label{eq:varianza_rho}
V \left[ \rho \right]:= \left\langle r^2 \right\rangle -\left\langle r \right\rangle^2= \frac{\eta^2 (\eta^2+2)-L^2 (L+1)^2}{4 Z^2}
\end{equation}
in position space \cite{ray:ajp88,dehesa:ijqc10}. In momentum space, we know \cite{dehesa:ijqc10,aptekarev:jpa10} that
\begin{equation}\label{eq:p_p2}
\left\langle p \right\rangle = \frac{2 Z}{\pi \eta} \quad \text{and} \quad \left\langle p^2 \right\rangle = \frac{ Z^2}{ \eta^2}
\end{equation}
so that the momentum variance
\begin{equation}\label{eq:varianza_gamma}
V \left[ \gamma \right]:= \left\langle p^2 \right\rangle -\left\langle p \right\rangle^2= \frac{Z^2}{\eta^2} \left(1-\frac{4}{\pi^2} \right) 
\end{equation}

Then, the use of Eqs. (\ref{eq:definicion_CR}), (\ref{eq:shannon_rho}) and (\ref{eq:varianza_rho}) allows us to find the value
\begin{equation}\label{eq:Ccr_posiciones}
C_{\text{CR}} \left[ \rho \right]:= F\left[ \rho \right] \times V \left[ \rho \right]= \frac{1}{\eta^3} \left(n- \left| m \right| \right) \left[\eta^2 (\eta^2+2)-L^2 (L+1) \right]
\end{equation}
for the Cram\'er-Rao complexity in position space. Similarly, from Eqs. (\ref{eq:definicion_CR}), (\ref{eq:shannon_gamma}) and (\ref{eq:varianza_gamma}) one obtains the value
\begin{equation}\label{eq:Ccr_momentos}
C_{\text{CR}} \left[ \gamma \right]:= F\left[ \gamma \right] \times V \left[ \gamma \right]= 2 \left( 1- \frac{4}{\pi^2} \right) \left[5\eta^2-3 L (L+1)- \left| m \right| (8\eta -6L-3)-1 \right]
\end{equation}
for the momentum Cram\'er-Rao complexity. Again here, the Cram\'er-Rao complexities do not depend on the nuclear charge $Z$ in both conjugated spaces. It is remarkable the explicit expressions of these complexity measures in terms of the hyperquantum numbers which characterize the $D$-dimensional hydrogenic orbital under consideration.\\

Finally, let us point out that this $D$-dimensional methodology of the two-component complexities has been recently applied \cite{dehesa:epjd09} to the standard or three-dimensional hydrogenic atoms, obtaining the full dependence of these quantities on the quantum numbers ($n,l,m$) which characterize the ground and the excited states of these realistic systems.


\section{LMC shape complexity of circular orbitals}
\label{sec:section3}

In this Section we apply the general expressions (\ref{eq:Clmc_posiciones}) and (\ref{eq:Clmc_momentos}) of the position and momentum LMC shape complexity, respectively, for a large class of stationary $D$-dimensional hydrogenic states, the circular states, which correspond to the electronic orbitals with the highest hyperangular momenta allowed within a given electronic manifold, i.e., the states with hyperangular quantum numbers $\displaystyle{\mu_i=n-1}$ for all $\displaystyle{i=1,\cdots,D-1}$. This class includes the ground state, which has the quantum numbers $n=1$ and $\displaystyle{\mu_i=0}$, $\displaystyle{\forall i=1,\cdots,D-1}$.\\

According to Eqs. (\ref{eq:rho_Dens}) and (\ref{eq:gamma_Dens}), the circular state (c.s.) have the expressions
\begin{equation}\label{eq:rho_circular}
\rho_{\text{c.s.}}(\vec{r})= \frac{2^{D+2-2n} Z^D  }{\pi^{\frac{D-1}{2}}(2n+D-3)^{D}
 \Gamma(n) \Gamma \left(n+\frac{D-1}{2} \right)  } e^{-\frac{r}{\lambda} } 
\left( \frac{r}{\lambda} \right)^{2n-2} \prod^{D-2}_{j=1} \left( \sin \theta_j \right)^{2n-2}
\end{equation}
and
\begin{equation}\label{eq:gamma_circular}
\gamma_{\text{c.s.}}(\vec{p})=\frac{2^{2n-2} (2n+D-3)^D \Gamma 
\left( n+\frac{D-1}{2}\right)}{Z^D \pi^{\frac{D+1}{2}} \Gamma(n)} 
\frac{(\eta p/Z)^{2n-2}}{(1+\frac{\eta^2 p^2}{Z^2})^{2n+D-1}} 
\prod^{D-2}_{j=1} \left( \sin \theta_j \right)^{2n-2}
\end{equation}
for the probability densities in position and momentum space, respectively.\\

Then, from Eqs. (\ref{eq:Drho1}) and (\ref{eq:Dgamma1}) one can calculate the explicit expressions for the position and momentum disequilibrium of any circular state as
\begin{equation}\label{eq:desequilibrio_rho_cs}
D \left[ \rho_{\text{c.s.}} \right]=\frac{Z^D \Gamma \left(n-\frac{1}{2} \right) 
\Gamma \left(2 n + \frac{D-3}{2} \right)}{2^{2n-2} \pi^{\frac{D}{2}} (2n+D-3)^D 
 \Gamma \left(n \right) \Gamma^2 \left(n+\frac{D-1}{2} \right)},
\end{equation}
and
\begin{equation}\label{eq:desequilibrio_gamma_cs}
D \left[ \gamma_{\text{c.s.}} \right]=\frac{2^{4n+D-4} (2n+D-3)^D 
\Gamma^2 \left( n+\frac{D-1}{2} \right) \Gamma \left(2n-1 \right) \Gamma
 \left( 2n+\frac{3D}{2} \right)}{Z^D \pi^{\frac{D+2}{2}} \Gamma^2 \left(n \right) \Gamma \left( 4n+2D-2 \right)}
\end{equation}
respectively. Moreover, working similarly from Eqs. (\ref{eq:shannon_rho}) and (\ref{eq:shannon_gamma}) we obtain the following values
\begin{align}\nonumber
S \left[\rho_{\text{c.s.}} \right]&= 2n+D-2 -(n-1) \left[ \psi (n) +
 \psi \left(n+\frac{D-1}{2} \right)  \right] -D \ln 2\\ \label{eq:shannon_rho_cs}
&\qquad+\ln \left[(2n+D-3)^D \pi^{\frac{D-1}{2}} \Gamma(n) 
\Gamma  \left(n+\frac{D-1}{2} \right)\right]- D \ln Z,
\end{align}
and
\begin{equation}\label{eq:shannon_gamma_cs}
S \left[\gamma_{\text{c.s.}} \right]= A(n,D)+\ln \left[ \frac{2^{D+1} Z^D 
\pi^{\frac{D+1}{2}} \Gamma(n) }{(2n+D-3)^D \Gamma  \left(n+\frac{D-1}{2} \right) }\right],
\end{equation}
for the position and momentum Shannon entropies, respectively, where the constant $\displaystyle{A(n,D)}$ is given by
\begin{align}\nonumber
A(n,D)=&\frac{2n+D-1}{2n+D-3}-\frac{D+1}{2n+D-2}-(n-1) \psi (n) \\\label{q:coefA_cs}
&\quad -\left(\frac{D+1}{2}\right) \psi \left(n+\frac{D-2}{2} \right) 
+\left(n+\frac{D-1}{2}\right) \psi \left(n+\frac{D-3}{2} \right).
\end{align}

Note that we have used the notation $\displaystyle{\psi(x)= \Gamma^{'}(x)/ \Gamma(x)}$ for the digamma function.\\

Then, from Eq. (\ref{eq:Clmc_posiciones}) one obtains the value
\begin{align}\nonumber
C_{\text{LMC}} \left[ \rho_{\text{cs}} \right]& = \frac{\Gamma  \left(n-\frac{1}{2} \right)
 \Gamma \left(2n+\frac{D-3}{2} \right)}{2^{2n+D-2} \pi^{1/2} 
\Gamma \left(n + \frac{D-1}{2} \right)}\\ \label{eq:Clmc_posiciones_cs}
& \quad \times \exp \left\lbrace 2n+D-2-(n-1) \left[ \psi(n) + \psi
 \left(n+ \frac{D-1}{2} \right) \right] \right\rbrace,
\end{align}
for the LMC shape complexity in position space. And from Eq. (\ref{eq:Clmc_momentos}) one finds the value
\begin{equation}\label{eq:Clmc_momentos_cs}
C_{\text{LMC}} \left[ \gamma_{\text{cs}} \right]=\frac{2^{4n+2D-3}\Gamma \left(n+
 \frac{D-1}{2} \right) \Gamma(2n-1) \Gamma \left(2 n + \frac{3D}{2} \right)}
{\pi^{1/2} \Gamma(n) \Gamma(4n+2D-2)} \exp \left[ A(n,D) \right]
\end{equation}
for the LMC shape complexity in momentum space.\\

For $n=1$ the last two expressions allow us to find the values
\begin{equation}\label{eq:Clmc_posiciones_gs}
C_{\text{LMC}} \left[ \rho_{\text{gs}} \right]= \left( \frac{e}{2} \right)^D
\end{equation}
and
\begin{align}\label{eq:Clmc_momentos_gs}
C_{\text{LMC}} \left[ \gamma_{\text{gs}} \right]& = \frac{2^D \Gamma \left( \frac{D+1}{2}\right)
 \Gamma \left(2+\frac{3 D}{2} \right) } {\pi^{1/2} \Gamma \left( 2D+2\right)} \nonumber \\
&\quad \times \exp \left\lbrace  (D+1) \left[ \psi \left( D+1 \right)-\psi \left( \frac{D+2}{2} \right)  \right]  \right\rbrace
\end{align}
for the position and momentum LMC shape complexities, respectively, of the ground state of the $D$-dimensional hydrogenic system. Let us point out that these values agree with the three-dimensional values previously found \cite{catalan:pre02}.\\

\begin{figure}[ht]
\begin{center}
\includegraphics[scale=0.3,angle=270]{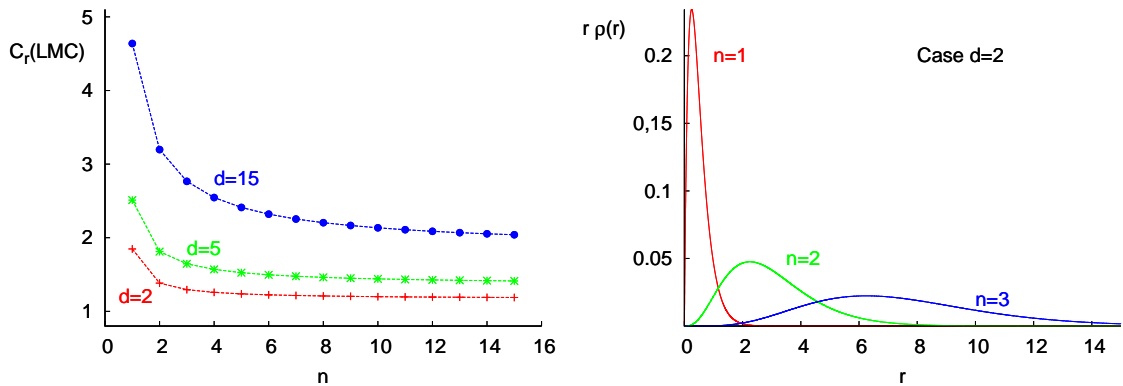}
\end{center}
\caption{(Left) Dependence of the position $LMC$ complexity of circular states on the 
principal quantum number $n$ for various dimensionalities. (Right)  Radial 
probability density in position space for various two-dimensional circular states.}
\label{fig:figura1}
\end{figure}

Finally, from the sake of completeness, we have shown in Figure \ref{fig:figura1} the dependence of the position LMC shape complexity of circular states with various dimensionalities on the principal quantum number $n$. Therein, we observe that the complexity decreases at all dimensionalities considered in the figure ($D=2,5$ and $15$) when $n$ is increasing.\\

A more detailed analysis shows that when the quantum number $n$ is increasing, the radial density $\rho_{\text{cs}}(\vec{r})$ behaves so that its maximum height decreases and its spreading increases at different rates in such a way that the total balance is the following: the larger $n$ is, the smaller is the complexity of the corresponding circular state. For further details and applications of LMC complexity, see \cite{lopezrosa:pa09b}.


%

\section{Conclusions}
\label{sec:section4}

In summary we have surveyed the theoretical methodology recently developed to compute the LMC shape, the Fisher-Shannon and the Cram\'er-Rao complexities of $D$-dimensional hydrogenic sytems in both position and momentum spaces. Then, we have illustrated it in the study of the LMC complexity of the circular hydrogenic states. For the sake of completeness, we would like to call the attention on the Fisher-R\'enyi and LMC-R\'enyi complexities \cite{lopezruiz:jmp09,antolin:cpl09,romera:irp09} whose analytical determination for hydrogenic systems is still lacking. This family of five two-component complexities provides a quite complete determination of the internal disorder of hydrogenic systems and, as shown by other contributions to this issue, other atomic and molecular systems from their electron distributions.

\section*{Acknowledgments}

We are very grateful for the partial financial support of the grants 
FIS2008-02380, FQM-2445, FQM-4643 and FQM-207 of the Junta de Andaluc\'ia and Ministerio de 
Ciencia e Innovaci\'on, Spain, EU.


\bibliography{bibliografia_01-07.bib}

\bibliographystyle{dcu}

\end{document}